# Nanoscale Phase Change Memory with Graphene Ribbon Electrodes

Ashkan Behnam[1], Feng Xiong[1,2], Andrea Cappelli[3], Ning C. Wang[1,2], Enrique A. Carrion[1], Sungduk Hong[1], Yuan Dai[1], Austin S. Lyons[1], Edmond K. Chow[1], Enrico Piccinini[4], Carlo Jacoboni[3], and Eric Pop[1,2,*]

[1]Department of Electrical and Computer Engineering & Micro and Nanotechnology Lab
University of Illinois at Urbana-Champaign, Urbana, IL 61801, USA

[2]Department of Electrical Engineering, Stanford University, Stanford, CA 94305, USA

[3]Department of Physics, Mathematics and Computer Science, University of Modena and Reggio Emilia, Via Campi 213/B, I-41125 Modena, Italy

[4]Department of Electrical, Electronic and Information Engineering "Guglielmo Marconi", University of Bologna, Viale Risorgimento 2, I-40136 Bologna, Italy

Phase change memory (PCM) devices are known to reduce in power consumption as the bit volume and contact area of their electrodes are scaled down. Here, we demonstrate two types of low-power PCM devices with lateral graphene ribbon electrodes: one in which the graphene is patterned into narrow nanoribbons and the other where the phase change material is patterned into nanoribbons. The sharp graphene "edge" contacts enable switching with threshold voltages as low as ~3 V, low programming currents (<1 μA SET, <10 μA RESET) and ON/OFF ratios >100. Large-scale fabrication with graphene grown by chemical vapor deposition also enables the study of heterogeneous integration and that of variability for such nanomaterials and devices.

*Corresponding author: epop@stanford.edu

Electrically-programmable phase change memories (PCMs) have captivated wide interest for applications in non-volatile memory[1] and reprogrammable circuits[2] due to low power operation[3-5], fast access times,[6] and high endurance.[7] Data in PCMs are stored by the large ratio (>$10^3$) in electrical resistance between the amorphous (OFF) and crystalline (ON) states of the material. A drawback of PCMs has been the traditionally high programming current (> 0.1 mA), which can be mitigated by reducing the cell volume and contact area, and by carefully engineering the electrical and thermal coupling between the phase change material and contacts.[1,8] The minimum energy required to switch PCM bits has been estimated to be as low as 1.2 aJ/nm$^3$ for thermally well-insulated, nanoscale memory bits.[1] In this context, nanomaterials such as carbon nanotubes (CNTs)[3-5] and graphene ribbons[9] are promising candidates for achieving small PCM contact area due to their atomically sharp edges and excellent operation at current densities required to program the PCM (~$10^7$ A/cm$^2$, while CNTs and graphene can carry ~$10^9$ A/cm$^2$, much higher than metals[10]). Indeed, CNTs and graphene have been successfully tested as electrodes in various types of non-volatile memory structures.[3-5,11,12,13] With recent advances in large-scale and low-cost fabrication of CNTs and graphene, they could also be used in transparent and flexible low-power electronics that often exhibit limited thermal budget.[12]

In this study, we use graphene ribbons as lateral "edge" electrodes to induce reversible phase change in small volumes of chalcogenide-based PCM, in this case $Ge_2Sb_2Te_5$ (GST). Chemical vapor deposition (CVD) has facilitated wafer-scale and low-cost growth of high quality graphene that can be transferred easily to various substrates and controllably patterned for device fabrication.[14] In our structures, large graphene sheets are transferred to $SiO_2$ substrates and then patterned into narrow interconnects in contact with GST, using lithography and dry etching techniques. Our designs allow control over the programming current and power of PCM devices, providing an excellent platform to study their scalability and performance using standard fabrication methods.

We developed two different structures to characterize the performance of PCM devices with graphene electrodes (Fig. 1). For both structures, first single-layer graphene sheets are grown on copper foils using the CVD technique. After chemically etching the copper foils, graphene sheets are transferred to 90 nm $SiO_2$ on highly doped Si (p+ type) substrates (Figs. S1-1 to S1-6 in Supplement[15]). Then the surface of the graphene is cleaned and the substrate is annealed. Details of the graphene growth and transfer process are provided in Section A of the Supplementary Material.[15] Raman spectroscopy, optical imaging and atomic force microscopy (AFM) analysis of the graphene surface suggests that most of the as-grown graphene is monolayer with average grain size >200 nm.[10] However, most PCM devices in this paper utilized few-layer graphene electrodes, obtained by repeating the growth, transfer and cleaning process of monolayer graphene 3 or 4 times on the same substrate.

For the first set of structures [Fig. 1(a)-(d)], graphene is patterned into nanoribbons (GNRs) contacting a wider GST material to form the PCM cell. First, large Ti/Au (0.5/30 nm) probing pads are defined by optical lithography and electron-beam (e-beam) evaporation [Fig. 1(a)]. The graphene under these contacts is removed by a 20-second $O_2$ plasma etch before metal deposition to facilitate better adhesion between contact pads and the $SiO_2$ substrate. This step is followed by creating smaller Pd/Au (30/30 nm) finger electrodes using e-beam lithography and evaporation. These electrodes are in contact with both the large probing pads and the graphene underneath [Figs. 1(b)-(d)]. Nanoribbons with small gaps are then patterned on the graphene by e-beam lithography and 3-10 nm of Al is deposited on the developed regions using e-beam evaporation to protect the graphene underneath (Fig. S1-7a). Part of the thin Al film oxidizes when the sample is removed from the evaporation chamber, and the $Al/AlO_x$ nanoribbons cover the graphene and stretch between finger electrodes after the e-beam resist lift-off (Fig. S1-8a).

Finally, a 20-second $O_2$ plasma etch removes all unprotected graphene, leaving GNR electrodes under the Al etch mask (Fig. S1-8a). The size of the nanogap between these GNR electrodes defines the nominal length of the PCM cell ($L_G$) and the GNR width defines the nominal PCM cell width ($W$). The protective Al layer is then chemically etched (Transene Al etch Type A) and ~10 nm of GST is deposited in the gap between the two GNR electrodes by e-beam lithography and sputtering, completing the PCM cell formation [Fig. 1(b)-(d) and Fig. S1-9a]. The electrical properties of ~10 nm thin GST films were characterized in our previous work,[16] showing >$10^3$ times change in resistivity from the amorphous to the crystalline state around 150 ºC. After GST deposition a protective $SiO_2$ layer (10 nm) is typically e-beam evaporated on the sample to improve the durability of the fabricated GNR-PCM structures. The lengths ($L$) and widths of the GNR electrodes are 0.5–1 μm and 30–400 nm, respectively. The nanogap length ($L_G$) is 30–100 nm. A group of control devices are also fabricated alongside the

GNR-PCM devices with similar structures. However, instead of GNRs narrow metallic (Ti/Au, 0.5/30 nm) fingers form the electrodes of the PCM cells. These metallic fingers are defined by e-beam lithography and deposited by e-beam evaporation ($W > 100$ nm).

For the second set of structures [Fig. 1(e)-(g)], wider graphene microribbons (GµR) are patterned as electrodes and the GST is deposited as a narrow nanoribbon instead of the larger window used for the GNR devices. As a result, the approximate PCM cell width ($W$) is defined by the lateral extent of the GST nanoribbons (GSTNR) that connect GµR electrodes [Figs. 1(e) and 1(g)]. Similar to GNR structures, the first step of GSTNR device fabrication is to define large Ti/Ni (0.5/30 nm) probing pads using optical lithography and e-beam evaporation [Fig. 1(a)]. Then, a gold-based shadow evaporation technique[17] is used, followed by $H_2$ plasma etching to controllably create a gap of 20–100 nm in the graphene between the probing pads [inset of Fig. 1(e), Fig. S1-7b and Section B in the Supplementary Material[15]]. The graphene is then patterned into microribbons (GµR with widths of 2–10 µm) using photolithography [Fig. 1(e)-(g)] and the rest of the graphene is removed by $O_2$ plasma etching (Fig. S1-8b). The last two steps are the definition (using e-beam lithography) and deposition of a GST nanoribbon ($W \sim 50$ nm, thickness 10 nm) across the gap in the GµR electrodes to form the PCM cell, and evaporation of the protective $SiO_2$ layer (10 nm) on top [Fig. 1(e)-(g) and Fig. S1-9b]. The active area of the GSTNR structures is defined by the width of the GSTNR ($W$) and the gap length ($L_G$).

Both GNR and GSTNR structures allow us to evaluate the scalability of the PCM devices with graphene electrodes. While GNRs are more desirable for the proper scaling of the cell structure, GSTNR structures facilitate more accurate extraction of device parameters such as contact resistance between GST and graphene due to the small dimensions of the GST nanoribbons and the well-defined structure of PCM device active area.

Figures 2 and 3 summarize the PCM cell operation results for all the structures. Figure 2(a) shows DC current sweeps and switching of a GNR device, demonstrating large resistance change between the crystalline (ON) and amorphous (OFF) states (examples for GSTNR structures are provided in Fig. S3).[15] In addition, devices switch almost instantaneously from OFF to ON state (SET operation). Low bias OFF and ON resistances ($R_{OFF}$ and $R_{ON}$ normalized with respect to $W$) are compared in Fig. 2(b). OFF/ON resistance ratios for most devices range between 10 and 1000 depicting successful switching across all types of structures, with a median value around ~100. Some of the best results belong to GSTNR devices with OFF/ON resistance ratios above $10^3$, close to the inherent resistance ratio of GST between its amorphous and FCC crystalline states.[18] These high OFF/ON ratios correspond to the well-defined and small PCM bit structure in GSTNR devices, suggesting that either most GST bridging the gap between graphene electrodes is being switched from the amorphous to crystalline states, or that a large crystalline pathway (within an amorphous matrix) connects the electrodes.[19]

The distribution of RESET current densities (applied to switch PCM bits back from crystalline to amorphous state) is presented in Fig. 2(c). Higher voltages (5 to 20 V) are applied using short (~100 ns) pulses to RESET the devices and current densities ($J_{RESET}$) are estimated by dividing the current through the nominal PCM bit cross-sectional areas. The distribution of $J_{RESET}$ for GSTNR devices is broad, yet the average (~ $2 \times 10^6$ A/cm$^2$) is comparable to previously reported values for devices with similar cross-sectional areas. For GNR and control devices, however, the distributions are narrower and the average values are on the lower range of those previously reported.[20] We relate the lower $J_{RESET}$ estimated for GNR and control devices to the smaller effective area of the PCM material that is being switched between crystalline and amorphous states (compared to the nominal cross sectional area). For GSTNR devices the effective and nominal switching areas are closer due to the small dimensions of the GST confined between electrodes, and therefore the estimated $J_{RESET}$ values are more accurate. However, due to the small PCM cell dimensions in GSTNR devices and the large variations in device parameters including $W$, the distribution of $J_{RESET}$ is also wider for these devices. $J_{RESET}$ values reported in Fig. 2(c) also depend on the applied RESET pulse voltages and therefore do not represent the absolute minimum values attainable.

Current-voltage characteristics [Fig. 2(a)] enable better understanding of the switching threshold parameters. Distributions of SET threshold fields ($F_T$) and current densities ($J_T$) are presented in Fig. 3. There is no significant correlation between $J_T$ and $F_T$ values but, as expected, $J_T$ values are significantly lower than $J_{RESET}$. $F_T$ values are calculated without subtracting the voltage drop at the graphene-GST contact and hence are not representative of the intrinsic threshold fields. Nevertheless, the obtained values are comparable with those reported for other lateral devices (e.g. $F_T \sim 0.6$ MV/cm for GST bridge devices with TiN contacts[21] and $F_T \sim 1$ MV/cm for CNT-contacted GST memory cells[3]). Figure 4(a) presents the scaling of $R_{ON}$ and $R_{OFF}$ with threshold voltage ($V_T$) for various GSTNR devices. In our previous studies with CNT electrodes[3,5] we observed a linear correlation between $R_{ON,OFF}$ and $V_T$, both scaling with the intrinsic length and resistance of the PCM bit. However, here we see non-negligible scatter among the data and no clear dependence between the PCM resistance and $V_T$. We attribute this to the variation in the access resistance at the GST-graphene interface, which is caused by defects induced in the graphene electrodes during the GST sputtering process (see Supplementary[15] Section D and Raman data shown in Fig. S4). Compared to devices in our previous studies with CNT electrodes,[3,5] the PCM devices with graphene electrodes require approximately an order of magnitude higher programming power, which is consistent with the larger contact area between the graphene edges and PCM. Nevertheless, graphene electrodes are transparent and flexible, and may be better suited for large-scale lithography-based device fabrication.

Figures 4(b) and 4(c) present examples of the retention and switching of our PCM devices. For these measurements ON and OFF switching is performed by applying 500 and 100 ns voltage pulses, respectively, with voltages ranging from 3 to 20 V depending on device type and dimensions. Resistance varies only slightly over time for cells programmed in both ON and OFF states [Fig. 4(b)]. However, these PCM bits tend to fail after several ON/OFF manual switching cycles, as shown in Fig. 4(c). The poor switching performance of these devices could be related to several factors including insufficient GST encapsulation (exposure to air over time) and poor interfaces between GST and graphene electrodes. We note that the switching is indeed occurring in the GST layer and not in the $SiO_2$ underneath the graphene nanogap since the applied electric field is significantly lower than that needed to induce switching in $SiO_2$.[12,13] We tested several GNR control devices before GST deposition and we observed no switching in the oxide. Nonetheless, we suggest that the reliability of future graphene-PCM devices could be improved by better encapsulation schemes (e.g. with insulators devoid of oxygen, like $Si_3N_4$), by improving the graphene-GST interface, by using vertical PCM devices perpendicular to the substrate, and by avoiding over-programming during RESET operation.

To better understand the state of the graphene-GST interface, we can also use our device structures to estimate the contact resistivity ($\rho_C$) between the graphene electrodes and GST in the ON and OFF states, as shown in Fig. 5. These contributions include the GST resistance (in the crystalline and amorphous states, respectively) and graphene electrode sheet and access resistances. Details of $\rho_C$ estimation are provided in Section E and Fig. S5 of the Supplementary Material.[15] Figure 5 shows that our graphene-contacted structures exhibit a wide range of contact resistivities due to both device-to-device process variations and estimation uncertainties. The best results, however, are comparable to measurements of metal-contacted PCM devices that have found $\rho_{C,OFF}$ between $7\times10^{-4}$ and $10^{-1}$ $\Omega\cdot cm^2$ and $\rho_{C,ON}$ between $2\times10^{-7}$ and $5\times10^{-6}$ $\Omega\cdot cm^2$, respectively.[22] (Details of our contact resistivity estimates are provided in Supplementary Section E and Table S1.) In addition, overall performance of graphene-based devices, particularly those with small dimensions, is comparable to or better than the control devices most likely due to the smaller effective contact area between GST and the atomically sharp graphene edges. These results suggest that graphene has the potential to be used as a contact material in non-volatile PCM devices. Surely, further process improvements would enable enhanced performance and large-scale fabrication of such non-volatile memory structures with graphene electrodes.

In conclusion, we have demonstrated lateral PCM devices with patterned graphene ribbon electrodes. The thin structure of these devices (with thin PCM and graphene layers) could make them appealing for flexible and transparent electronics that have strict low-power requirements, as well as for integration with conventional CMOS substrates. Although the power consumption of these devices is approximately an order of magnitude greater than those with CNT electrodes[3,5] (consistent with the larger contact area between graphene edges and PCM), programming currents are nevertheless in the single µA range, threshold voltages are as low as ~3 V, and median OFF/ON ratios are ~100. However, the variability and reliability of these devices must be improved by decreasing the variation in process parameters, and by controlling the confinement and quality of the GST material and its interface with the substrate and contacts.

This work was in part supported by the Office of Naval Research (ONR) Young Investigator Award N00014-10-1-0853, the Air Force Office of Scientific Research (AFOSR) grant FA9550-14-1-0251, and by Systems on Nanoscale Information Fabrics (SONIC), one of six SRC STARnet Centers sponsored by MARCO and DARPA.

# References

1       S. Raoux, F. Xiong, M. Wuttig, and E. Pop,  MRS Bulletin **39**, 703 (2014).
2       S.M. Yoon, S.W. Jung, S.Y. Lee, Y.S. Park, and B.G. Yu,  IEEE Electron Device Letters **30**, 371 (2009).
3       F. Xiong, A.D. Liao, D. Estrada, and E. Pop,  Science **332**, 568 (2011).
4       J.L. Liang, R.G.D. Jeyasingh, H.Y. Chen, and H.S.P. Wong,  Ieee Transactions on Electron Devices **59**, 1155 (2012).
5       F. Xiong, M.H. Bae, Y. Dai, A.D. Liao, A. Behnam, E.A. Carrion, S. Hong, D. Ielmini, and E. Pop,  Nano Letters **13**, 464 (2013).
6       D. Loke, T.H. Lee, W.J. Wang, L.P. Shi, R. Zhao, Y.C. Yeo, T.C. Chong, and S.R. Elliott, Science **336**, 1566 (2012).
7       C. Chieh-Fang, A. Schrott, M.H. Lee, S. Raoux, Y.H. Shih, M. Breitwisch, F.H. Baumann, E.K. Lai, T.M. Shaw, P. Flaitz, R. Cheek, E.A. Joseph, S.H. Chen, B. Rajendran, H.L. Lung, and C. Lam,  IEEE International Memory Workshop (IMW), DOI: 10.1109/IMW.2009.5090589 (2009).
8       K.L. Grosse, F. Xiong, S. Hong, W.P. King, and E. Pop,  Applied Physics Letters **102**, 193503 (2013);       H.L. Ma, X.F. Wang, J.Y. Zhang, X.D. Wang, C.X. Hu, X. Yang, Y.C. Fu, X.G. Chen, Z.T. Song, S.L. Feng, A. Ji, and F.H. Yang,  Applied Physics Letters **99**, 173107 (2011);    Y.C. Fu, X.F. Wang, J.Y. Zhang, X.D. Wang, C. Chang, H.L. Ma, K.F. Cheng, X.G. Chen, Z.T. Song, S.L. Feng, A. Ji, and F.H. Yang,  Applied Physics a-Materials Science & Processing **110**, 173 (2013);         I.R. Chen and E. Pop, IEEE Transactions on Electron Devices **56**, 1523 (2009).
9       H. Tian, H.Y. Chen, B. Gao, S.M. Yu, J.L. Liang, Y. Yang, D. Xie, J.F. Kang, T.L. Ren, Y.G. Zhang, and H.S.P. Wong,  Nano Letters **13**, 651 (2013).
10      A. Behnam, A.S. Lyons, M.H. Bae, E.K. Chow, S. Islam, C.M. Neumann, and E. Pop, Nano Letters **12**, 4424 (2012).
11      F. Xiong, A. Liao, and E. Pop,  Applied Physics Letters **95**, 243103 (2009);        C.L. Tsai, F. Xiong, E. Pop, and M. Shim,  Acs Nano **7**, 5360 (2013);    S. Lee, J. Sohn, H.-Y. Chen, and H.-S.P. Wong,  arXiv, 1502.02675 (2015).
12      J. Yao, J. Lin, Y.H. Dai, G.D. Ruan, Z. Yan, L. Li, L. Zhong, D. Natelson, and J.M. Tour, Nature Communications **3**, 1101 (2012).
13      A. Shindome, Y. Doioka, N. Beppu, S. Oda, and K. Uchida,  Japanese Journal of Applied Physics **52**, 04CN05 (2013).
14      S. Bae, H. Kim, Y. Lee, X.F. Xu, J.S. Park, Y. Zheng, J. Balakrishnan, T. Lei, H.R. Kim, Y.I. Song, Y.J. Kim, K.S. Kim, B. Ozyilmaz, J.H. Ahn, B.H. Hong, and S. Iijima,  Nature Nanotechnology **5**, 574 (2010);        M. Bianchi, E. Guerriero, M. Fiocco, R. Alberti, L. Polloni, A. Behnam, E.A. Carrion, E. Pop, and R. Sordan,  Nanoscale **7**, 8076 (2015).
15       See supplementary material at http://dx.doi.org/XXX for the details of graphene growth and transfer process, shadow evaportaion technique for nanogap creation, extraction of contact resistivity between graphene and GST and for further examples of current-voltage characteristics for PCM devices. .
16      K.L. Grosse, E. Pop, and W.P. King,  Journal of Applied Physics **116**, 124508 (2014).
17      Y. Naitoh, M. Horikawa, H. Abe, and T. Shimizu,  Nanotechnology **17**, 5669 (2006).
18      M. Lankhorst, B. Ketelaars, and R. Wolters,  Nature Materials **4**, 347 (2005).

[19] A. Cappelli, E. Piccinini, F. Xiong, A. Behnam, R. Brunetti, M. Rudan, E. Pop, and C. Jacoboni, Applied Physics Letters **103**, 083503 (2013); E. Piccinini, A. Cappelli, F. Xiong, A. Behnam, F. Buscemi, R. Brunetti, M. Rudan, E. Pop, and C. Jacoboni, in *IEEE Intl. Electron Devices Meeting (IEDM)* (2013), p. 601.

[20] H.S.P. Wong, S. Raoux, S. Kim, J.L. Liang, J.P. Reifenberg, B. Rajendran, M. Asheghi, and K.E. Goodson, Proc. of IEEE **98**, 2201 (2010); A. Pirovano, A.L. Lacaita, A. Benvenuti, F. Pellizzer, S. Hudgens, and R. Bez, in *IEEE Intl. Electron Devices Meeting (IEDM)* (2003), pp. 699.

[21] D. Krebs, S. Raoux, C.T. Rettner, G.W. Burr, M. Salinga, and M. Wuttig, Applied Physics Letters **95**, 082101 (2009).

[22] D. Roy, M.A.A. in't Zandt, and R.A.M. Wolters, IEEE Electron Device Letters **31**, 1293 (2010); E.K. Chua, R. Zhao, L.P. Shi, T.C. Chong, T.E. Schlesinger, and J.A. Bain, Applied Physics Letters **101**, 012107 (2012); D. Roy, M.A.A. in't Zandt, R.A.M. Wolters, C.E. Timmering, and J.H. Klootwijk, in *Non-Volatile Memory Technology Symposium (NVMTS)* (2009), pp. 12; S.D. Savransky and I.V. Karpov, in *Material Research Society (MRS)* (2008), Vol. 1072, pp. G06; D.L. Kencke, I.V. Karpov, B.G. Johnson, L. Sean Jong, K. DerChang, S.J. Hudgens, J.P. Reifenberg, S.D. Savransky, Z. Jingyan, M.D. Giles, and G. Spadini, in *IEEE Intl. Electron Devices Meeting (IEDM)* (2007), pp. 323; J. Lee, E. Bozorg-Grayeli, S. Kim, M. Asheghi, H.S.P. Wong, and K.E. Goodson, Applied Physics Letters **102**, 191911 (2013).

**Figures**

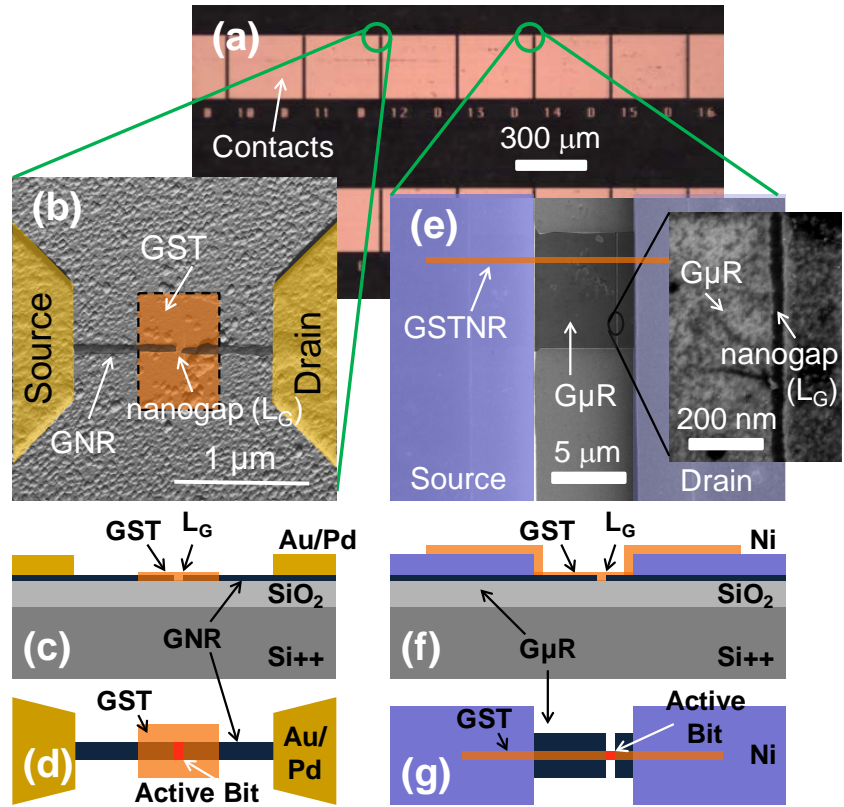

**FIG. 1.** Design of lateral PCM devices with graphene electrodes. **(a)** Optical image of the large Ti/Au or Ti/Ni contacts fabricated on Si/SiO$_2$ substrates. **(b), (c), (d)** Atomic force microscopy (AFM) image, cross section and top view cartoon of a lateral graphene nanoribbon (GNR) PCM device. In the AFM image $L$ = 1.5 μm, $W$ = 50 nm, and nanogap length ($L_G$) = 50 nm. GST thickness is 10 nm and GST window is 1×0.7 μm. **(e), (f), (g)** Scanning electron microscope (SEM) image, cross section and top cartoons of a lateral GST nanoribbon (GSTNR) PCM cell across graphene microribbon (GμR) electrodes. The average nanogap size between the GμR electrodes is $L_G$ ≈ 50 nm, as shown in the inset.

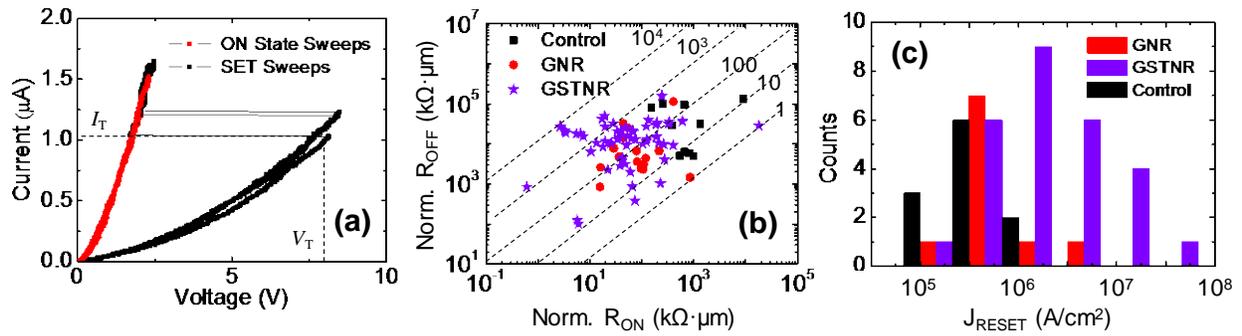

**FIG. 2.** **(a)** Memory switching of a GNR PCM device with $L_G \sim 70$ nm and $W \sim 30$ nm. **(b)** OFF vs. ON resistance values ($R_{OFF}$ vs. $R_{ON}$) normalized by $W$ for GNR, GSTNR and control devices with metal electrodes. Dashed lines show the $R_{OFF}/R_{ON}$ ratio contours, as labeled from 1 to $10^4$. **(c)** Distribution of RESET current density ($J_{RESET}$) for GNR, GSTNR and control devices.

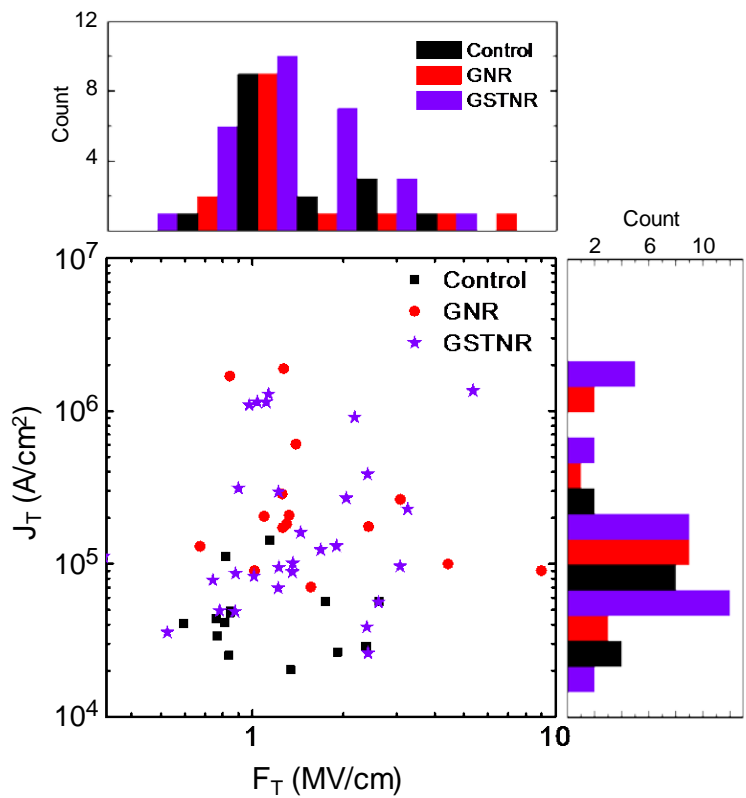

**FIG. 3.** Threshold current density ($J_T$) vs. threshold field ($F_T$) for various GSTNR, GNR, and control PCM devices (main panel). The panels on the right and top show the distributions of $J_T$ and $F_T$, respectively.

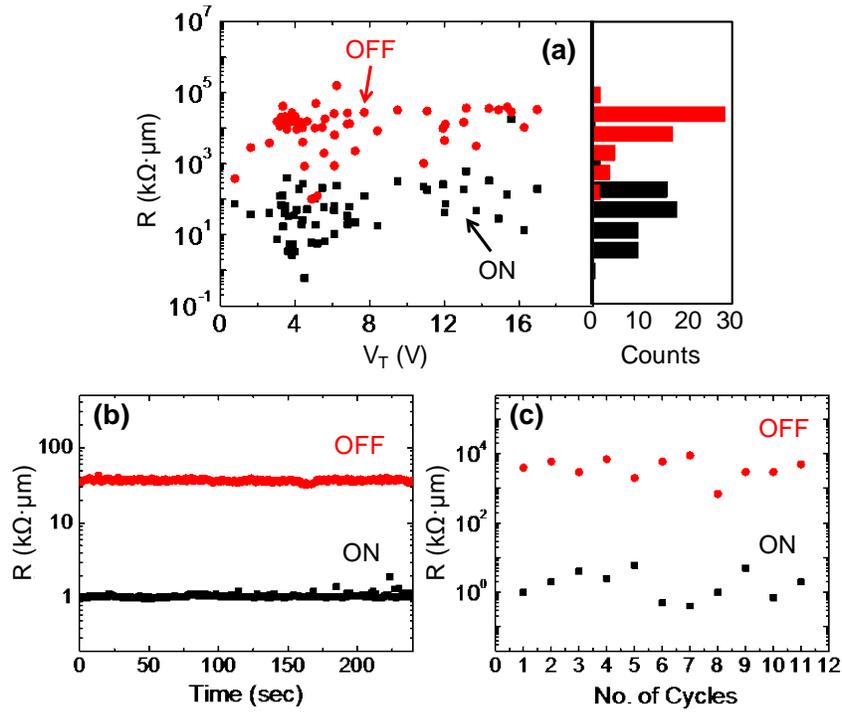

**FIG. 4.** **(a)** $R_{ON}$ and $R_{OFF}$ normalized by the bit width vs. device threshold voltage ($V_T$) for GSTNR devices. The OFF/ON ratio of all measured devices was summarized in Fig. 2(b). **(b), (c)** Endurance test results. **(b)** ON and OFF resistances for a GNR device (with $W = 400$ nm and $L_G \sim 70$ nm) are stable under a constant 1 V readout. This is equivalent to over $2 \times 10^9$ read operations with 100 ns pulses. **(c)** Resistance variation after consecutive OFF and ON cycles for a GNR device with $W = 400$ nm and $L_G \sim 70$ nm.

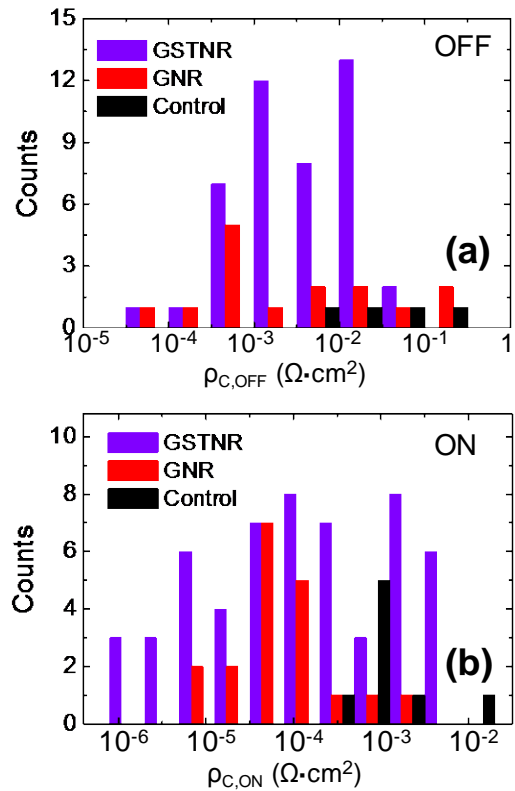

**FIG. 5.** Distribution of estimated contact resistivity ($\rho_C$) between GST and graphene in **(a)** OFF state and **(b)** ON state for GSTNR, GNR, and control PCM devices. Devices with different dimensions and nanogap sizes are considered.

# Supplementary Information

## Nanoscale Phase Change Memory with Graphene Ribbon Electrodes

Ashkan Behnam[1,*], Feng Xiong[1,2], Andrea Cappelli[3], Ning C. Wang[1,2], Enrique A. Carrion[1], Sungduk Hong[1], Yuan Dai[1], Austin S. Lyons[1], Edmond K. Chow[1], Enrico Piccinini[4], Carlo Jacoboni[3], and Eric Pop[1,2,*]

[1]Department of Electrical and Computer Engineering & Micro and Nanotechnology Lab
University of Illinois at Urbana-Champaign, Urbana, IL 61801, USA

[2]Department of Electrical Engineering, Stanford University, Stanford, CA 94305, USA

[3]Department of Physics, Mathematics and Computer Science, University of Modena and Reggio Emilia, Via Campi 213/B, I-41125 Modena, Italy

[4]Department of Electrical, Electronic and Information Engineering "Guglielmo Marconi", University of Bologna, Viale Risorgimento 2, I-40136 Bologna, Italy

[*]Corresponding author: epop@stanford.edu

## A. Graphene Fabrication and Transfer

Graphene growth by chemical vapor deposition (CVD) is performed by flowing $CH_4$ and Ar gases (100 sccm and 1000 sccm respectively) at 1000 °C and 500 mTorr chamber pressure, which results primarily in monolayer graphene growth on both sides[1] of the Cu foil (Fig. S1-1). One side of the graphene is protected by two layers of polymethyl methacrylate (PMMA) (with molecular weights of 495K and 950K) while the graphene on the other side of the foil is removed with a few-second 20 sccm $O_2$ plasma reactive ion etch (RIE) process (Fig. S1-2). The Cu foil is then etched overnight in aqueous $FeCl_3$ (Transene CE-100), leaving the graphene supported by PMMA floating on the surface of the solution (Fig. S1-4). The PMMA/graphene film is transferred via a glass slide to a modified SC-2 (20:1:1 $H_2O/H_2O_2/HCl$) bath, and then to two separate deionized water baths. Next, the film is transferred to the $SiO_2$ (90 nm ± 5 nm) on Si substrate (p+ doped, <5 mΩ·cm resistivity) and left for about an hour to dry (Fig. S1-5). The PMMA coating on the surface is removed using a 1:1 mixture of methylene chloride and methanol, followed by a 1-hour $Ar/H_2$ anneal at 400 °C to remove PMMA and other organic residue (Fig. S1-6). Transfer and anneal processes are repeated 3-4 times to obtain clean multi-layer graphene coverage. Substrates are inspected by optical microscope, Raman spectroscopy

2and atomic force microscopy (AFM) techniques (after the first transfer). The monolayer nature of the graphene and its relatively good quality are confirmed by these observations.[2]

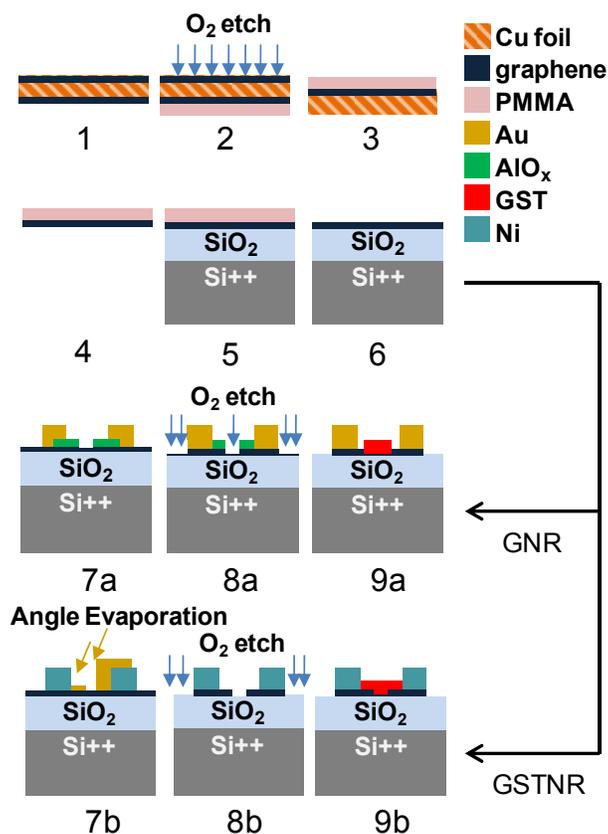

**Fig. S1.** Cross-section schematic of the fabrication process for GNR and GSTNR devices.

## B. Shadow Evaporation Technique for Creating Nanogaps

Gold (Au) shadow evaporation is used for creating the nanogap in GST nanoribbon (GSTNR) structures. First, a lithography step defines a large window in the resist material with an edge in between the Ni electrodes and in parallel with them. This window is filled with 35-50 nm thick Au using electron-beam (e-beam) evaporation and the rest of the resist is lifted off [Fig. S2(a)]. Then a second Au e-beam evaporation step is performed (thickness of ~15 nm) but this time at an angle $\theta$ ($20 < \theta < 45$ degrees) with respect to the line perpendicular to the surface of the sample (evaporation direction is perpendicular to the edge of the Au window in between Ni contacts). This blanket (shadow) evaporation step (with no lithography) leaves a small line gap (with no Au evaporated) in the middle of the Ni electrodes and parallel to their edge [Fig. S2(b) and Fig. S1-7b]. Flexibility in choosing the thickness of the first Au layer and the second Au layer deposition angle $\theta$ allows for controlling the gap length. After shadow evaporation the graphene in the gap region is etched using $H_2$ plasma (the rest protected by Au) and then all the Au is chemically removed (Transene Au etch) without significant damage to the Ni electrodes and graphene underneath.

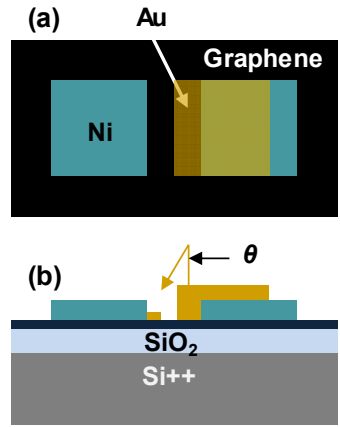

**Fig. S2.** (a) Top view schematic of the GSTNR sample after the deposition of first Au layer for shadow evaporation. (b) Cross-section of the sample in second Au deposition step at an angle $\theta$ with respect to the line perpendicular to the surface of the sample.

## C. Current-Voltage Sweeps of PCM devices

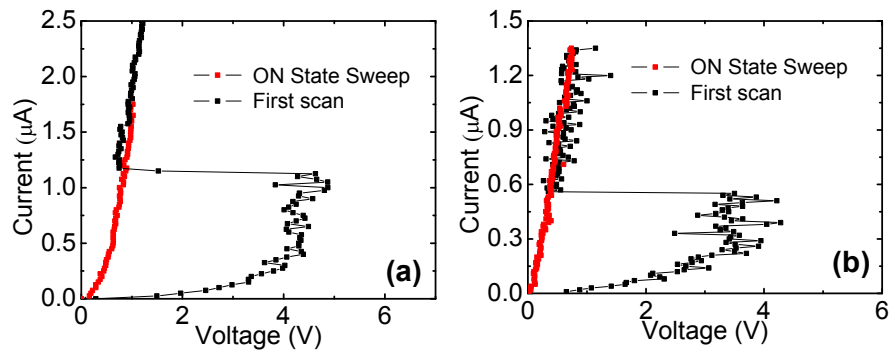

**Fig. S3.** Current sweeps in ON/OFF states of GSTNR memory devices. (a) Nanogap $L_G \sim 50$ nm and $W \sim 60$ nm. (b) $L_G \sim 50$ nm and $W \sim 30$ nm.

## D. Raman Spectroscopy of Graphene Before and After GST Deposition

Raman spectroscopy of graphene devices before and after GST deposition by sputtering shows a notable increase in the D/G peak ratio after the sputtering process, indicating graphene damage from the GST atom bombardment during sputtering [Fig. S4(a)]. The D Raman peak of graphene is well-known to be an indicator of graphene lattice damage and imperfections.[1,2]

Electrical measurements of graphene with and without GST coverage reveal that increasing the Ar pressure during sputtering (from 3 to 10 mTorr) reduces the defects induced in graphene from the GST sputtering process [Fig. S4(b)]. This occurs ostensibly because the GST atoms reaching the graphene surface have lower average kinetic energy when the Ar pressure is higher, as they undergo more collisions before they are deposited onto the graphene. Nevertheless, the GST deposition process on graphene is far from "perfect" and appears in large part responsible for the variability and contact resistance seen at the GST-graphene interfaces. Further improvements of this deposition process could yield much improved graphene-PCM devices in the future.

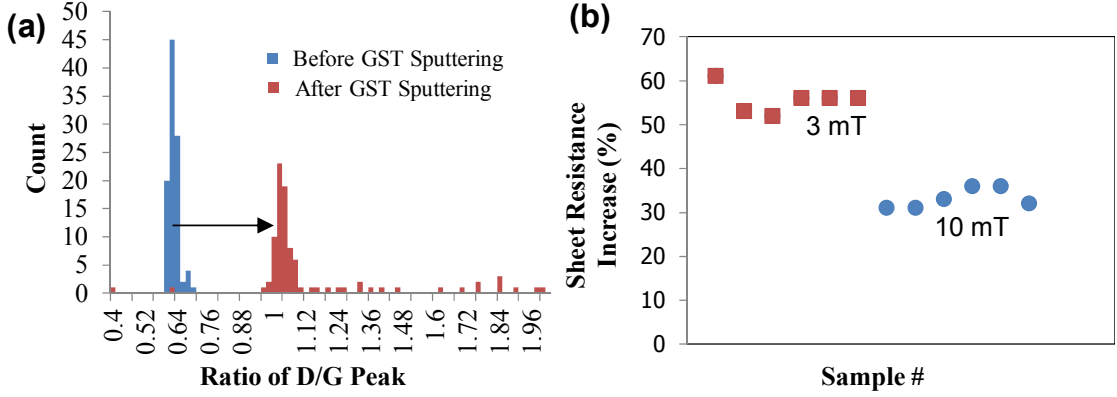

**Fig. S4.** (a) Distribution of the ratio of D/G peak intensity in Raman spectroscopy of over 100 graphene samples before and after GST sputtering. Increase in the D/G ratio suggests damage to graphene after GST deposition. (b) Increasing the Ar pressure during sputtering helps mitigate the damage. As we increase the Ar pressure from 3 to 10 mTorr during GST sputtering, the sheet resistance of the graphene under GST increases by almost ~30% less across several samples.

### E. Graphene-GST Contact Resistivity Extraction and Other Studies Findings

The measured resistance of the PCM devices ($R_{Total}$) consists of several components including the resistance of metal electrodes ($R_M$), the contact resistance between graphene and metal electrodes ($R_{M-G}$), the resistance of the graphene ribbons ($R_G$), the contact resistance between graphene ribbons and GST ($R_{G-GST}$), and finally the resistance of the GST bit ($R_{GST}$) (Fig. S5):

$$R_{Total\_on,off} = 2R_M + 2R_{M-G} + 2R_G + 2R_{G-GSTon,off} + R_{GSTon,off} \qquad (S1)$$

The first three elements are independently estimated from four-point probe measurements on the metals (for $R_M$) and measurements on graphene nanoribbons with different dimensions but no gaps (for $R_G$ and $R_{M-G}$). $R_M$ is small compared to $R_G$ and $R_{M-G}$ (due to the large size of the metallic pads) and can be neglected in calculations.

The two last elements ($R_{G-GST}$ and $R_{GST}$) add up to a significant portion of the total resistance and are the only elements that significantly depend on the state of the PCM cell (ON vs. OFF). The resistivity of GST in the OFF and ON states is taken to be $10^2$ Ω·cm and $10^{-2}$ Ω·cm, respectively.[3] Based on these values, $R_{G-GST}$ is estimated in the OFF and ON mode. Then the contact resistivity $\rho_{Con,off}$ in the ON/OFF states is calculated from:[4]

$$R_{GSTon,off} = \frac{\rho_{Con,off}}{L_T Z} \coth\left[\frac{L_C}{L_T}\right] \qquad (S2)$$

where $Z$ is the width of the contact region between graphene and GST, $L_C$ is the length of the overlap region (Fig. S5) and $L_T = (\rho_{Con,off} / R_S)^{1/2}$ is the transfer length defined as the position inside the contact region ($x = L_T$) in the transport direction at which the electric potential becomes a fraction $1/e$ of its value at the edge of the contact ($x = 0$). $R_S$ is the sheet resistance of graphene. Equation S2 could be further simplified in the limit that $L_C$ becomes significantly larger or smaller than $L_T$, but since in our devices $L_C$ is quite comparable to $L_T$ (especially for GNR devices $L_C$ < 500 nm), the non-simplified form of the equation is solved recursively to obtain $\rho_{Con,off}$ values. For metal control devices, instead, equation S1 is further simplified due to the fewer number of elements contributing to the total resistance values. The results of the contact resistance calculations are given in Fig. 5 of the main text.

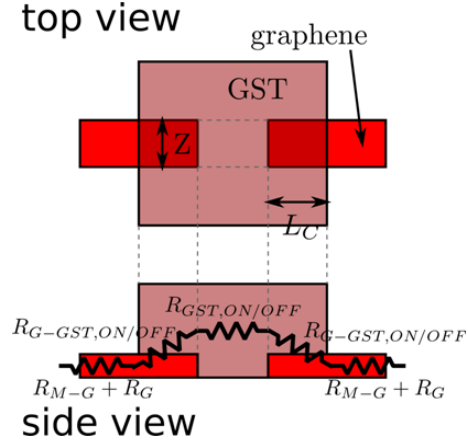

**Fig. S5.** Different components of the total resistance in a GNR PCM cell.

| Materials | $\rho_{C,OFF}$ ($\Omega cm^2$) | $\rho_{C,ON}$ ($\Omega cm^2$) | Type of Measurement | Ref. |
|---|---|---|---|---|
| GST-TiW | $7 \times 10^{-4}$ | $5 \times 10^{-6}$ @200°C | Circular Transmission Line Method (C-TLM) | [5] |
| GST-TiW | $9.5 \times 10^{-2}$ | ~ $4 \times 10^{-6}$ @175°C<br>~ $2 \times 10^{-7}$ @350°C | Cross Bridge Kelvin Resistance (CBKR) | [6] |
| GST – Metal-Nitrides | – | ~ $3.5 \times 10^{-7}$ | Transmission Line Method (TLM) | [7] |
| GST-Metal | – | ~ $2.5 \times 10^{-7}$ | TLM | [8] |
| GST-TiN | – | ~ $8 \times 10^{-7}$ @200°C<br>~ $7 \times 10^{-8}$ @300°C | CBKR and TLM | [9] |
| GST-graphene | ~ $5 \times 10^{-4}$ - $5 \times 10^{-2}$ | ~ $10^{-6}$ - $10^{-3}$ | Estimated | This work |

**Table S1.** Comparison of PCM contact resistivity estimated in this work (with graphene electrodes, bottom row) with other metallic contact materials reported in the literature.[5-9]

**Supplementary References**

1. X. S. Li, W. W. Cai, J. H. An, S. Kim, J. Nah, D. X. Yang, R. Piner, A. Velamakanni, I. Jung, E. Tutuc, S. K. Banerjee, L. Colombo, and R. S. Ruoff, *Science* **324**, 1312 (2009).

2. A. Behnam, A. S. Lyons, M-H. Bae, E. K. Chow, S. Islam, C. M. Neumann, and E. Pop, *Nano Lett.* **12**, 4424 (2012).

3. M. Lankhorst, B. Ketelaars, and R. Wolters, *Nat. Mater.* **4**, 347 (2005).

4. D. K Schroder, *Semiconductor material and device characterization* (Wiley, 2006).

5. D. Roy, M. A. A. in't Zandt, R. A. M. Wolters, C. E. Timmering, and J. H. Klootwijk, presented at the Non-Volatile Memory Technology Symposium (NVMTS), pp. 12-15 (2009).

6. D. Roy, M. A. A. in't Zandt, and R. A. M. Wolters, *IEEE Electron Dev. Lett.* **31** (11), 1293 (2010).

7. S. D. Savransky and I. V. Karpov, in *Material Research Society Spring Meeting Proceedinngs*, Vol. 1072, pp. G06-09 (2008).

8. D. L. Kencke, I. V. Karpov, B. G. Johnson, Lee Sean Jong, Kau DerChang, S. J. Hudgens, John P. Reifenberg, S. D. Savransky, Zhang Jingyan, M. D. Giles, and G. Spadini, *IEEE Electron Dev. Meet.*, pp. 323-326 (2007).

9. J. Lee, E. Bozorg-Grayeli, S. Kim, M. Asheghi, H.-S. P. Wong, and K. E. Goodson, *Appl. Phys. Lett.* **102**, 191911 (2013).